\documentclass[epj,twocolumn]{svjour_mod}
\usepackage{amsmath} 
\usepackage{amssymb}
\usepackage{graphicx,epsfig}
\usepackage[mathscr]{eucal}
\usepackage{color}
\usepackage{longtable}
\usepackage[english]{babel}
\usepackage{booktabs}
\usepackage{setspace}
\usepackage{xspace}
\usepackage{multirow}
\usepackage{cite}
\usepackage{scalefnt,pstricks}

\newcommand\as{\alpha_s} 
\newcommand{\beq}{\begin{equation}}
\newcommand{\eeq}{\end{equation}}

\usepackage[colorlinks=true,allcolors={blue!70!black}]{hyperref}
\usepackage{footnotebackref}

\begin{document} 

\author{
Rikkert Frederix \inst{1}
\and
Paolo Torrielli \inst{2}
}
\date{}
\title{A new way of reducing negative weights in MC@NLO}
\institute{
Division of Particle and Nuclear Physics, Department of Physics,
Lund University, Box 118, SE-221 00 Lund, Sweden
\and
Dipartimento di Fisica, Universit\`a di Torino and INFN, Sezione di Torino,
I-10125 Torino, Italy
\\
\email{rikkert.frederix@hep.lu.se, paolo.torrielli@unito.it}
}

\abstract{
We introduce a new technique, that we dub {\it Born spreading}, aimed at reducing the number of negative-weight $\mathbb S$ events in the MC@NLO matching of NLO calculations with parton-shower simulations. We show that such a technique, based on a re-distribution of Born matrix elements in the radiative phase space, achieves a sizeable reduction of negative-weight events at little computational cost. The method does not induce any biases in physical distributions. 
}
\maketitle

\section{Introduction}

Modern high-energy particle phenomenology requires Monte Carlo simulations featuring steadily increasing computational costs, both in terms of running times and of computing resources~\cite{HEPSoftwareFoundation:2020daq,CERN-LHCC-2022-005,Software:2815292}.
The bulk of the simulations carried out by the experimental collaborations are ones where next-to-leading order (NLO) predictions in QCD are matched with parton-shower event generators. Since NLO cross sections are in general not positive-definite, it is not uncommon that some of the events generated by means of such simulations are associated to a negative weight.

As far as resources are concerned, the presence of negative weights may represent a serious bottleneck. The computational cost of a simulation, for a given target accuracy (Monte Carlo error), scales as $1/(1-2f)^2$, with $f$ being the total fraction of negative-weight events in the sample~\cite{Frederix:2020trv}. Such an issue may be severe for the MC@NLO matching scheme~\cite{Frixione:2002ik}, where particularly complex LHC processes may produce up to $O(30-40\%)$ negative-weight events, whereas alternative matching schemes~\cite{Nason:2004rx,Frixione:2007vw,Jadach:2015mza,Nason:2021xke} are less affected.

Various techniques have been proposed aiming at the reduction of negative weights, either modifying the event generation~\cite{Frederix:2020trv,Danziger:2021xvr}, or as a post-processing step on existing event samples~\cite{Andersen:2020sjs,Nachman:2020fff,Andersen:2021mvw,Andersen:2023cku}.
In the context of MC@NLO, and in particular in its implementation within the MadGraph5\_aMC@NLO code~\cite{Alwall:2014hca} (abbreviated as MG5\_aMC henceforth), negative weights have two distinct sources. Negative $\mathbb H$ events have a physical origin in the way the MC@NLO matching is defined, i.e.~they depend on whether the radiation probability, as predicted by the parton shower one is interfacing to, is bigger than the NLO real matrix element for the considered process. As such, the fraction of these negative-weight events can be reduced only with a modification of the matching prescription, e.g.~the one that has been proposed in~\cite{Frederix:2020trv} and dubbed MC@NLO-$\Delta$ matching. 

On the other hand, negative $\mathbb S$ events are largely caused by the fact that, although having Born kinematics, the short-distance cross section associated with them has support in the radiative phase space. Such a cross section is not positive-definite locally in the radiative phase space, resulting in negative-weight events after the unweighting procedure. This is usually tackled by means of the folding procedure~\cite{Nason:2007vt,Frixione:2007nu,Alioli:2010xd,Frederix:2020trv}, where at a given Born phase-space point one averages over many triplets (\emph{folds}) of variables associated to the radiated particle, thereby efficiently compensating local  $\mathbb S$-integrand negativities before the unweighting step is taken. Although this is a systematic working solution, it typically entails a significant runtime increase before a reduction of negative weights is obtained, roughly proportional to the number of folds, so that the final effective gain in terms of computational costs is not always a priori obvious.

In this paper we propose a novel alternative strategy, that we dub \emph{Born spreading}, to reduce negative $\mathbb S$ events in MG5\_aMC at very little CPU cost with respect to the standard MC@NLO implementation. We present the ideas behind the method in Section \ref{sec:BS}, we detail its implementation in Section \ref{sec:impl} and its performances with respect to folding in Section \ref{sec:res}. In Section \ref{sec:ext} we discuss a number of possible extensions of the method, and draw our conclusions in Section \ref{sec:conc}.

\section{Born spreading}
\label{sec:BS}
The $\mathbb S$-event generating functional in MC@NLO can be schematically written as
\beq
\label{eq:Sevts}
{\cal F}^{(\mathbb S)}
\, = \,
\int \Big[
\frac{B(\Phi_B)}{\int d\Phi_r} +
\frac{V(\Phi_B)}{\int d\Phi_r} +
K_{\mathrm{MC}}(\Phi_B,\Phi_r)
\Big]d\Phi_r \times
{\cal F}_{\mathrm{MC}}^{(B)}
\, ,
\eeq
where $d\Phi_B$ and $d\Phi_r$ are the Born and radiative phase-space measures, respectively (the former including Bjorken fractions in case of hadronic collisions, as well as flux factors). $B(\Phi_B)$ is the Born matrix-element squared, whereas $K_{\mathrm{MC}}(\Phi_B,\Phi_r)$ is the so-called Monte Carlo counterterm, i.e.~the $O(\as)$ emission probability as approximated by the shower (a local subtraction of infrared and collinear singularities from this term, e.g.~by means of the FKS procedure~\cite{Frixione:1995ms}, is understood). The term labelled as $V(\Phi_B)$ is the finite part of the renormalised virtual contribution, including remnants of the integrated (FKS) subtraction terms, as well as mass factorisation finite contributions in case of hadronic collisions. Finally, ${\cal F}_{\mathrm{MC}}^{(B)}$ is the standard generating functional of the parton shower, where the label $(B)$ indicates that the partonic multiplicity the shower evolution starts from is the Born-level one. Equation \eqref{eq:Sevts} makes it clear that, although $\mathbb S$-event kinematics is Born-like, the support of the integrand in square brackets is in the full Born+radiative phase space.

In Equation \eqref{eq:Sevts}, the origin of negative {$\mathbb S$}-event contributions is twofold\footnote{There is, potentially, also a third small contribution to the negative weights coming from non-positive parton density functions (PDFs).}. The first one is due to the FKS subtraction in $K_{\mathrm{MC}}(\Phi_B,\Phi_r)$: even though the subtraction cancels when integrated over $\Phi_r$, locally it might over-compensate, resulting in a negative contribution. The second source is that for certain $\Phi_B$ configurations, the virtual contribution, $V(\Phi_B)$, can be negative. Both these sources can be reduced by the following procedure.

Since the dependence of the Born contribution $B(\Phi_B)$ upon the radiative variables $\Phi_r$ is immaterial, the expression in Equation \eqref{eq:Sevts} can be equivalently rewritten as
\begin{multline}
\label{eq:Sevts2}
{\cal F}^{(\mathbb S)}
\, = \,
\int \Big[
\frac{B(\Phi_B) \, F(\Phi_r)}{\int F(\Phi_r) \, d\Phi_r} +
\frac{V(\Phi_B)}{\int d\Phi_r} +\\
K_{\mathrm{MC}}(\Phi_B,\Phi_r)
\Big]d\Phi_r \times
{\cal F}_{\mathrm{MC}}^{(B)}
\, ,
\end{multline}
where an arbitrary integrable {\it spreading function} $F(\Phi_r)$ has been introduced as a Born multiplier. It is clear that the functional form of $F(\Phi_r)$ does not affect physical distributions stemming from the $\mathbb S$-event generating functional, as those just depend on Born-level kinematics, and the spreading function is normalised so as to preserve the Born weight locally in $\Phi_B$. Therefore, one can leverage the ample freedom in the definition of $F(\Phi_r)$, and find one that minimises the presence of negative weights in the $\mathbb S$ events.
The idea is that of defining $F(\Phi_r)$ so that the Born contribution, now non-uniformly distributed over $\Phi_r$, is accumulated more where the rest of the $\mathbb S$-event integrand is more negative, as to rescue as much as possible of the negatively-contributing phase space.

If we define
\beq
W(\Phi_B,\Phi_r)
\, = \,
\Big[
\frac{V(\Phi_B)}{\int d\Phi_r} +
K_{\mathrm{MC}}(\Phi_B,\Phi_r)
\Big]\frac{1}{B(\Phi_B)}\, ,
\eeq
namely the $\mathbb S$-event contribution of non-Born origin with the Born divided out, a spreading function that satisfies the above requirements is
\beq
\label{eq:F Phir}
F(\Phi_r)
\, = \,
\max\Big[0,-\int W(\Phi_B,\Phi_r) \, d\Phi_B \, \Big]
\, .
\eeq
We use this as our baseline choice to study the Born-spreading performances throughout the paper.

We conclude this Section by mentioning a subtlety in the Born-spreading procedure that specifically arises in the context of the MC@NLO matching. In MC@NLO one needs to assign to each generated event a shower starting scale, representing the phase-space boundary for subsequent shower activity. Such a scale can be chosen arbitrarily to a large extent, with the sole constraint that the Monte Carlo counterterms used in the definition of the short-distance MC@NLO cross section must feature the very same boundary, in order not to introduce double counting at $O(\as)$. As the contributions to ${\cal F}^{(\mathbb S)}$ in Equation (\ref{eq:Sevts}) have support in the radiative phase space, the starting scale assigned to $\mathbb S$ events may in principle depend on $\Phi_r$. Were this the case, any spreading function $F(\Phi_r)\neq 1$ would cause a distortion in the starting-scale distribution and, in turn, a bias in the prediction for physical observables through showering effects. We stress that this bias would not represent a higher-order effect, namely it would spoil the formal $O(\as)$ accuracy of the prediction.
Two main strategies can be adopted to correct for this feature. On the one hand, one can encode $F(\Phi_r)$ in the definition of the Monte Carlo conuterterms: however, on top of requiring a thorough validation of the latter, this solution would also modify the functional form of $W(\Phi_B,\Phi_r)$ and in turn of the spreading function itself, resulting in a circular approach. A simple alternative solution is to assign to $\mathbb S$ events a starting scale that just depends on $\Phi_B$: this option is not only formally allowed, but also more natural from the physical viewpoint, as it potentially reduces spurious correlations between radiative and non-radiative configurations. We have opted for this solution in our implementation of Born spreading in the context of MG5\_aMC, which has entailed minimal modifications to the default scale assignment for events in the Monte Carlo dead zones.

\section{Implementation}
\label{sec:impl}
Our implementation of the Born-spreading idea in MG5\_aMC builds upon the default integration and event-generation strategies of the latter, summarised in the following. A preliminary stage (dubbed {\it step-0}) is executed solely for the integration routines to iteratively adapt the multi-dimensional grids to the shape of the integrands at hand, while the integration results are discarded. Based on such grids, in a subsequent {\it step-1} the proper integration is performed, together with the evaluation of the upper bounds necessary for event unweighting. Finally, {\it step-2} deals with the generation of the unweighted (up to a sign) event sample.

If Born spreading is active, step-0 is initially run as usual, with the $\mathbb S$-integrand defined as in Equation (\ref{eq:Sevts}); after the integration grids are set up, the code turns to sample the spreading function $F(\Phi_r)$ in the radiative phase space. The latter is parametrised in terms of the dimensionless FKS variables $\xi$, $y$, $\phi$ \cite{Frixione:1995ms}, related to the energy, polar and azimuthal angle of the radiated parton. The azimuthal modulation of the integrands is discarded at all stages of the Born-spreading procedure, which is expected to have little numerical effect, as $\phi$ is not associated to any physical phase-space singularities. Then a two-dimensional $N_{\xi} \times N_{y}$ grid (by default $40 \times 40$) is defined, and the code throws $N_{\rm spread}$ random points (by default $10^6$) to sample the spreading function\footnote{The virtual component in the spreading function is evaluated only approximately (using $\tilde{V}_k$ instead of $V$, as defined in Ref.~\cite{Alwall:2014hca}), i.e.~no loop matrix-element routine is actually called in the spreading procedure.}. In each of the $N_{\xi} \times N_{y}$ bins, the piecewise value of the spreading function $F_{ij}$ (with $i \, (j) = 1,...,N_{\xi \, (y)}$) is obtained upon averaging over all of the sampled underlying Born kinematics. At this point, the $\mathbb S$-integrand is redefined as in Equation (\ref{eq:Sevts2}) (with the formal replacement $F(\Phi_r)\to F_{ij}$), and another instance of step-0 is run, in order to let the integration grids re-adapt to the newly-defined integrand. Finally, once the new grids have been obtained, the subsequent integration and event-generation steps proceeds as in the default code.

From the above discussion, one can anticipate that in presence of Born spreading the time spent by the code in the step-0 phase will be significantly larger than in the default operational mode: this is mainly driven by the sampling of the spreading function, and to a lesser extent by the subsequent grid setup. However, commonly step-0 is (by far) the fastest step in the whole integration and event-generation procedure, hence a time penalty at this stage is not particularly worrisome; moreover, such a penalty just depends on the considered process, and does not scale with the number of events produced, nor with the accuracy required for the integration result, thereby representing a mere constant time offset. Conversely, the subsequent step-1 and step-2 are expected to perform by construction as well as in the default code: this represents a considerable potential advantage in terms of runtimes with respect to folding, which was mentioned above to scale linearly with the number of employed folds.

\section{Results}
\label{sec:res}
In this Section we present results for the Born-spreading procedure applied to several LHC processes, covering a variety of physical situations including colour-singlet production ($pp\to e^+e^-$, and $pp\to H$), processes with jets at Born level ($pp\to W^+j$), and processes with massive coloured final states ($pp\to t\bar t$, $pp\to W^+t\bar t$, and $pp\to Hb\bar b$).
Our simulations refer to the LHC operated at 13 TeV, with the following Standard Model parameters: $m_t = 173$ GeV, $m_W = 80.385$ GeV, $\Gamma_W = 2.047600$ GeV, $m_H = 125$ GeV, $m_Z = 91.188$ GeV, $\Gamma_Z = 2.441404$ GeV, $G_F = 1.16639 \cdot 10^{-5}$ GeV$^2$, $\alpha = 1/132.507$. We employ the central replica of the NNPDF2.3 PDF set \cite{Ball:2012cx}, which sets the strong coupling value as $\alpha_s(m_Z) = 0.118$. The central values of renormalisation and factorisation scales are chosen to be $\mu_{R,F} = \frac12\sum_i(m_i^2 + p_{T,i}^2)^{1/2}$, with the sum running over all final-state Born-level particles.
All processes are simulated in the five-flavour scheme, with the exception of $pp\to Hb\bar b$, where a bottom mass $m_b=4.7$ GeV is assumed. For neutral Drell-Yan $pp\to e^+e^-$ we require a threshold for the lepton-pair invariant mass, $m_{e^+e^-}\geq 30$ GeV, while for $pp\to W^+j$ we impose a $p_T\geq 10$ GeV generation cut on the hardest jet in the event, where jets are reconstructed with FASTJET \cite{Cacciari:2011ma} using an $R=0.7$ $k_T$ algorithm.

In Table \ref{tab:results}, Born spreading is compared against two folding setups, with $n_\xi \times n_y \times n_\phi = 2\times2\times1$ and $4\times4\times1$ folds, respectively. Results of the default MG5\_aMC baseline (formally corresponding to a $1\times1\times1$ folding), are also displayed for reference.
The comparison focuses on the reduction of negative-weight $\mathbb S$ events with respect to the $1\times1\times1$ baseline, as well as on time performances
relevant to the generation of 1M events per process on a desktop machine. Runtimes are broken up in their different integration and event-generation stages. We have extensively verified that, upon addressing the subtlety presented at the end of Section \ref{sec:BS}, physical distributions obtained with Born spreading are statistically identical to those obtained with the default MG5\_aMC code and with folding, hence we refrain from showing the corresponding plots.

  \begin{table*}
\begin{center}
    
  \begin{tabular}{lcccc}
    \toprule
& step-0 (s)    & step-1 (s) & step-2 (s)  & negative \\
& (grid setup) & (integration) & (generation)  & $\mathbb S$ weights\\[2pt]
\midrule
$\boxed{pp\to e^+e^-}$     &     &     &        &       \\[4pt]
default                                  &   1 &  14 & 147 & 7.1\% \\[1pt]
$2 \times 2 \times 1$ folding            &   1 &  33 & 258 & 2.1\% \\[1pt]
$4 \times 4 \times 1$ folding            &   1 & 114 & 781 & 1.8\% \\[1pt]
\blue{Born spreading}   & \blue{113} &  \blue{30} & \blue{189}  & \blue{2.0\%} \\[2pt]
\midrule
$\boxed{pp\to H}$          &     &     &           &        \\[4pt]
default                    &   1 & 121 &  187 & 10.6\% \\[1pt]
$2 \times 2 \times 1$ folding            &   1 & 115 &  399 &  2.7\% \\[1pt]
$4 \times 4 \times 1$ folding            &   1 & 228 & 1190 &  0.6\% \\[1pt]
\blue{Born spreading}   &  \blue{82} & \blue{122} &  \blue{203} &  \blue{1.1\%} \\[2pt]
\midrule
$\boxed{pp\to t\bar t}$    &     &      &            &       \\[4pt]
default                &   2 &  132 &  455 &   8.6\% \\[1pt]
$2 \times 2 \times 1$ folding            &   2 &  262 & 1005 &  2.2\% \\[1pt]
$4 \times 4 \times 1$ folding            &   2 & 1092 & 3189 &  1.2\% \\[1pt]
\blue{Born spreading}   & \blue{199} &  \blue{137} &  \blue{448} & \blue{2.1\%} \\[2pt]
\midrule
$\boxed{pp\to W^+t\bar t}$ &     &      &            &       \\[4pt]
default                &   5 &  346 & 1511 & 4.2\% \\[1pt]
$2 \times 2 \times 1$ folding            &   2 &  661 & 2938 & 2.2\% \\[1pt]
$4 \times 4 \times 1$ folding            &   2 & 2605 &10020 & 1.7\% \\[1pt]
\blue{Born spreading}   & \blue{202} &  \blue{741} &  \blue{2138}  & \blue{2.6\%} \\[2pt]
\midrule
$\boxed{pp\to W^+j}$       &     &      &            &       \\[4pt]
default                &  10 &  604 & 2013 & 24.2\% \\[1pt]
$2 \times 2 \times 1$ folding            &  10 & 1265 & 5160 & 13.2\% \\[1pt]
$4 \times 4 \times 1$ folding            &   7 & 2803 &16020 & 9.0\% \\[1pt]
\blue{Born spreading}   & \blue{355} &  \blue{645} &  \blue{2226} & \blue{18.8\%} \\[2pt]
\midrule
$\boxed{pp\to H b \bar b}$ &     &      &            &       \\[4pt]
default                &  77 & 1311 &19440 & 27.3\% \\[1pt]
$2 \times 2 \times 1$ folding            &  39 & 4320 &16380 & 22.4\% \\[1pt]
$4 \times 4 \times 1$ folding            &  48 &17220 &34260 & 20.9\% \\[1pt]
\blue{Born spreading}   &  \blue{578}   & \blue{1263} &  \blue{20760} & \blue{24.7\%} \\[2pt]
\bottomrule
\end{tabular}
\caption{Runtimes and fraction of negative-weight $\mathbb S$ events for various LHC processes with default MG5\_aMC code, two folding setups, and Born spreading.}
\label{tab:results}
\end{center}
\end{table*}

As far as the first four listed processes are concerned, Born spreading proves an excellent compromise between speed and negative-weight reduction. The latter is substantial in comparison with the default code, with residual negative fraction at the level of $O(2\%)$. Correspondingly, as anticipated, Born spreading induces a significant runtime increase only in the grid-setup phase, which in any case is reasonably fast and, importantly, does not scale with the number of generated events; time performances remain comparable with the baseline at all subsequent stages, which represents a strong benefit of the Born-spreading procedure.
For these four processes, Born spreading is better than or comparable with $2\times2\times1$ folding in terms of negative-weight reduction, and is significantly faster. $4\times4\times1$ folding is systematically more efficient than Born spreading in reducing negative $\mathbb S$ weights, however it typically takes a factor $3-5$ longer. We argue that once the residual fraction of negative weights $f$ is at the level of few percent, reducing it further is barely justified, since in that case the computational-cost reduction scales linearly with $f$, while inducing a time penalty that scales with the number of events. In this respect, Born spreading represents an interesting alternative to folding.

As for processes with light QCD final states, $pp\to W^+j$ and $pp\to H b\bar b$, in general we note a sizeable fraction of negative $\mathbb S$ weights. Born spreading is able to systematically reduce their impact with respect to the default MG5\_aMC, maintaining comparable runtimes. In this case already $ 2 \times 2 \times 1$ folding outperforms Born spreading as for negative-weight reduction, and runtimes are only moderately increased at this level\footnote{Notably, step-2 in $pp\to H b\bar b$ with $2\times2\times1$ folding is even faster than the default: most of the integration time in this case is spent in the evaluation of the virtual contribution, which is not affected by folding. Hence the enhanced stability of the integrand achieved with folding is sufficient to guarantee smaller numerical error, and an increased unweighting efficiency.}. Being the fraction of negative weights still considerably large, the actual benefits of folding with respect to Born spreading have to be carefully assessed taking into account a realistic number of generated events, as well as time and CPU cost spent in the showering and detector-simulation phase.

\section{Possible extensions of the method}
\label{sec:ext}
The Born-spreading strategy introduced in this paper aims at alleviating the impact of negative $\mathbb S$ weights in MC@NLO simulations. Its goals are the same as the well-established folding procedure, with which it can be naturally compared.
On one hand, folding guarantees a progressive reduction of negative weights which scales with the number of employed folds. This means that in principle a mimimum fraction of negative weights can be reached. However, the resources needed to achieve a significant reduction may well outweigh the advantages of the reduction itself, which requires case-by-case assessment.
On the other hand, the Born-spreading technique on paper has a more limited scope than folding: being based on a re-distribution of the Born contribution in the radiative phase space, it does not fully capture how the distribution of negative weights is correlated with the Born phase-space variables. In this sense, once a spreading function has been defined, the procedure is not parametrically improvable. Nevertheless, the strategy is computationally so much cheaper than folding to warrant consideration as the new default operational mode in MG5\_aMC, especially for the production of large event samples. We note incidentally that the negative-weight events in the POWHEG approach \cite{Nason:2004rx} have the same origin as the negative $\mathbb S$ weights in MC@NLO, whence Born spreading could prove beneficial with that approach as well.

Although the Born-spreading results shown in Section \ref{sec:res} look encouraging, we stress that the analysis of the method is still at a preliminary stage, and much is to be done to fully exploit its potential. The first immediate direction would be that of optimising the involved parameters ($N_\xi$, $N_y$, $N_{\rm spread}$), in order to potentially reduce the time offset in step-0 without degrading the negative-weight reduction. Furthermore, the functional form of the spreading function itself should be more carefully assessed: if the one defined in Equation (\ref{eq:F Phir}) is quite natural, we notice that it just exploits the region in which the no-Born $\mathbb S$ cross section is locally negative. One could for instance explore spreading functions that erode
the positive integrand in order to fill more efficiently the negative region, achieving a further compensation of negative weights. Furthermore, the ample freedom in the choice of spreading function makes this problem an ideal test ground for neural-network algorithms, which would then have to learn the optimal spreading function minimising a loss function related to the fraction of negative weights. This method could also be expanded to capture some of the correlations between negative weights and Born variables that are lost in the current implementation of spreading.
Another way to extend the method is to include the virtual contribution in the spreading. It can either be spread together with the Born contribution, or, a separate spreading function can be defined for the virtual corrections. For the latter case, the Born and virtual spreading functions would be in competition and have to be determined with some iterative procedure, or learning algorithm. 
Moreover, Born spreading is also compatible with folding, namely one could envisage to act with spreading in step-0, and subsequently fold the spread sample, thereby reducing the number of required folds for a given target negative fraction. Finally, Born spreading achieves a reduction of negative weights which is naturally complementary to that of MC@NLO-$\Delta$, and their joint effect should be thoroughly assessed in realistic event generation.

\section{Conclusions}
\label{sec:conc}
We have presented a new method, dubbed \emph{Born spreading}, for the reduction of negative-weight $\mathbb S$ events in the MC@NLO matching formalism. The method is based on the consideration that the Born contribution, usually integrated along with all the other components of the $\mathbb S$ integrand, by its own nature just depends on non-radiative variables, hence it can be re-distributed (\emph{spread}) arbitrarily in the radiative phase space. This freedom is exploited to find a spreading function that maximally reduces the fraction of negative $\mathbb S$ weights.

We have described the implementation of this idea in the context of MadGraph5\_aMC@NLO, and characterised its performances for a variety of physical processes at the LHC in terms of running times and negative-weight reduction. A detailed comparison has been performed both against the default MadGraph5\_aMC@NLO mode, and against the folding method. Born spreading is particularly efficient in achieving moderate to sizeable (but not extreme) negative-weight reductions at negligible extra CPU cost, at variance with folding, which can achieve more consistent reductions, but typically requiring deployment of significant computing resources.

We have finally indicated several possible ways to extend the Born-spreading idea, in the hope of optimising its performances and of making it an established valuable tool for the reduction of negative-weight events.

\section*{Acknowledgements}
RF has been partially supported by the Swedish Research Council contract
numbers 2016-05996 and 2020-04423.
PT has been partially supported by the Italian Ministry of University
and Research (MUR) through grant PRIN 2022BCXSW9 and by Compagnia di
San Paolo through grant TORP\_S1921\_EX-POST\_21\_01.

\bibliographystyle{spphys}
\bibliography{BSpread}

\begin{thebibliography}{10}
\providecommand{\url}[1]{{#1}}
\providecommand{\urlprefix}{URL }
\expandafter\ifx\csname urlstyle\endcsname\relax
  \providecommand{\doi}[1]{DOI \discretionary{}{}{}#1}\else
  \providecommand{\doi}{DOI \discretionary{}{}{}\begingroup
  \urlstyle{rm}\Url}\fi

\bibitem{HEPSoftwareFoundation:2020daq}
T.~Aarrestad, et~al., in \emph{{Snowmass 2021}}, ed. by P.~Canal, et~al.
  (2020).
\newblock \doi{10.5281/zenodo.4009114}

\bibitem{CERN-LHCC-2022-005}
{ATLAS Software and Computing HL-LHC Roadmap}.
\newblock Tech. rep., CERN, Geneva (2022).
\newblock \urlprefix\url{https://cds.cern.ch/record/2802918}

\bibitem{Software:2815292}
{CMS Phase-2 Computing Model: Update Document}.
\newblock Tech. rep., CERN, Geneva (2022).
\newblock \urlprefix\url{https://cds.cern.ch/record/2815292}

\bibitem{Frederix:2020trv}
R.~Frederix, S.~Frixione, S.~Prestel, P.~Torrielli, JHEP \textbf{07}, 238
  (2020).
\newblock \doi{10.1007/JHEP07(2020)238}

\bibitem{Frixione:2002ik}
S.~Frixione, B.R. Webber, JHEP \textbf{06}, 029 (2002).
\newblock \doi{10.1088/1126-6708/2002/06/029}

\bibitem{Nason:2004rx}
P.~Nason, JHEP \textbf{11}, 040 (2004).
\newblock \doi{10.1088/1126-6708/2004/11/040}

\bibitem{Frixione:2007vw}
S.~Frixione, P.~Nason, C.~Oleari, JHEP \textbf{11}, 070 (2007).
\newblock \doi{10.1088/1126-6708/2007/11/070}

\bibitem{Jadach:2015mza}
S.~Jadach, W.~P\l{}aczek, S.~Sapeta, A.~Si\'odmok, M.~Skrzypek, JHEP
  \textbf{10}, 052 (2015).
\newblock \doi{10.1007/JHEP10(2015)052}

\bibitem{Nason:2021xke}
P.~Nason, G.P. Salam, JHEP \textbf{01}, 067 (2022).
\newblock \doi{10.1007/JHEP01(2022)067}

\bibitem{Danziger:2021xvr}
K.~Danziger, S.~H\"oche, F.~Siegert, hep-ph/2110.15211

\bibitem{Andersen:2020sjs}
J.R. Andersen, C.~G\"utschow, A.~Maier, S.~Prestel, Eur. Phys. J. C
  \textbf{80}(11), 1007 (2020).
\newblock \doi{10.1140/epjc/s10052-020-08548-w}

\bibitem{Nachman:2020fff}
B.~Nachman, J.~Thaler, Phys. Rev. D \textbf{102}(7), 076004 (2020).
\newblock \doi{10.1103/PhysRevD.102.076004}

\bibitem{Andersen:2021mvw}
J.R. Andersen, A.~Maier, Eur. Phys. J. C \textbf{82}(5), 433 (2022).
\newblock \doi{10.1140/epjc/s10052-022-10372-3}

\bibitem{Andersen:2023cku}
J.R. Andersen, A.~Maier, D.~Ma\^\i{}tre, Eur. Phys. J. C \textbf{83}(9), 835
  (2023).
\newblock \doi{10.1140/epjc/s10052-023-11905-0}

\bibitem{Alwall:2014hca}
J.~Alwall, R.~Frederix, S.~Frixione, V.~Hirschi, F.~Maltoni, O.~Mattelaer, H.S.
  Shao, T.~Stelzer, P.~Torrielli, M.~Zaro, JHEP \textbf{07}, 079 (2014).
\newblock \doi{10.1007/JHEP07(2014)079}

\bibitem{Nason:2007vt}
P.~Nason, {arXiv 0709.2085 [hep-ph]}  (2007)

\bibitem{Frixione:2007nu}
S.~Frixione, P.~Nason, G.~Ridolfi, hep-ph/0707.3081  (2007)

\bibitem{Alioli:2010xd}
S.~Alioli, P.~Nason, C.~Oleari, E.~Re, JHEP \textbf{06}, 043 (2010).
\newblock \doi{10.1007/JHEP06(2010)043}

\bibitem{Frixione:1995ms}
S.~Frixione, Z.~Kunszt, A.~Signer, Nucl. Phys. B \textbf{467}, 399 (1996).
\newblock \doi{10.1016/0550-3213(96)00110-1}

\bibitem{Ball:2012cx}
R.D. Ball, et~al., Nucl. Phys. B \textbf{867}, 244 (2013).
\newblock \doi{10.1016/j.nuclphysb.2012.10.003}

\bibitem{Cacciari:2011ma}
M.~Cacciari, G.P. Salam, G.~Soyez, Eur. Phys. J. C \textbf{72}, 1896 (2012).
\newblock \doi{10.1140/epjc/s10052-012-1896-2}

\end{thebibliography}

\end{document}